\documentclass[10pt,prl,twocolumn,preprintnumbers,amsmath,amssymb, superscriptaddress]{revtex4}
\usepackage{graphicx}% Include figure files
\usepackage{dcolumn}% Align table columns on decimal point
\usepackage{bm}% bold math
\usepackage{epsfig}
\usepackage{subfigure}
\makeatletter
\newenvironment{figurehere}
  {\def\@captype{figure}}
  {}

\begin{document}
\title{A Massively Parallel Time Domain Phase Field Model for Multi-domain Ferroelectric Device Simulation }

%%
%\author{Khalid Ashraf, Sayeef Salahuddin}
\author{Khalid Ashraf}
\email{kashraf@eecs.berkeley.edu}
%\affiliation{Dept. of Electrical Engineering and Computer Sciences,
%University of California, Berkeley, CA 94720}
\author{Sayeef Salahuddin}
\affiliation{Dept. of Electrical Engineering and Computer Sciences,
University of California, Berkeley, CA 94720}

\date{\today}

\begin{abstract}
 In this work, we report a massively parallel and time domain implementation of the 3D phase field model that can reach beyond micron scale and consider for arbitrary electrical and mechanical boundary conditions. The first part of the paper describes the theory and the numerical implementation of the model. A mixed-mode approach of  finite difference (FD) and finite element (FEM) grid has been used for calculating the nonlocal electrostatic and elastic interactions respectively. All the local and non-local interactions are shown to scale linearly up to thousands of processors. This massive parallelization allows us to directly compare our results with multiple experiments at the same size scale. The second part of the paper presents results of ferroelectric switching in devices based on the multi-ferroic BiFeO$_{3}$(BFO). We have particularly emphasized the importance of charge driven domain growth and the effect of electrical boundary conditions that explain the temporal evolution of ferroelectric domains observed in  recent experiments.  We also predict a mechanism of controlling domain size in the multi-domain ferroelectric switching that could be useful for practical applications. 
%The implication of the presence of this high AFM moment is discussed in the context of achieving deterministic electric field driven magnetic moment switching. 
%\noindent
%\\\\{\em Keywords}: Exchange bias, Spin Glass, AFM moment density, Bismuth Ferrite,
%azimuthal hysteresis.
\end{abstract}

\maketitle
\newpage
\section{INTRODUCTION}
\label{sec:INTRO}

Thin film devices incorporating ferro-electric and multi-ferroic materials have attracted substantial research effort worldwide \cite{eerenstein_nature06}. Understanding switching dynamics in these multi-domain ferroelectric films influenced by arbitrary electrostatic and mechanical boundary condition remains to be a significant challenge. The origin of the difficulty lies in the coupling of multiple order parameters in these materials and the spatial asymmetry introduced by the domain walls. The necessity of being self consistent between various competing energies coming from chemical, electrostatic and elastic origin makes the temporal evolution of polarization a numerically stiff problem. Also ferroelectric domain walls are typically of  the order of nm size whereas the whole pattern forms over micron sizes. This disparate length scales associated with ferroelectric domain walls and domains themselves necessitates a large degree of freedom to be simulated in order to achieve physically reasonable results. Under experimental conditions, the non-linear switching behavior makes it very difficult to design and characterize devices. Faced by these obstacles, it remains a significant challenge to make direct comparison of a simulation result with an experimental observation and also pursue rational device design using computational simulation. In this paper, we report a significant step forward by extending the  capability of conventional phase field models \cite{lqchen_JACS08}, extensively used for ferroelectric materials,  up to the micron scale where experiments are typically performed. The distinct features of the model are that it can simulate structures up to micron size in 3D, take arbitrary electrical and mechanical boundary conditions and achieve linear scaling in calculating all the local and non-local electrical, mechanical and chemical interactions. Reaching the micron scale simulation grid allows us to make direct comparison with experimental observations in arbitrary device structures such as that shown in Fig. 1(a). 

In the following section, we describe the physical equations, numerical implementation and performance. Using this model, we simulated the lateral switching on various surfaces of the multiferroic material BFO. We make direct comparison with multiple experiments recently reported on various surfaces of BFO. Finally, we propose a method to control domain size during polarization switching that could be crucial for device applications using this material. 

\section{THEORY}
\label{sec:theory}
A phase field model describes the thermodynamic free energy of a system in terms of a continuous field variable. All the participating energies of the system is described as a function of this order parameter. For example, for ferroelectrics, the polarization is a suitable order parameter for describing the thermodynamic energy of the system. A continuum approximation for describing the spatiotemporal  variation of the polarization facilitates describing the low energy dynamics of the system. Phase field model has been applied to understand ferroelectric domain switching \cite{chowdhury_APL07,rao_APL07,sohJACE06,zhangAM05,ahlualiaPRB05,WangAM2004}, strain effects \cite{VaithyanathanJAP06,tenne_Science06,Li_PRB06,Li_APL06,Li_JAP05,chowdhury_JACE05,tenne_PRB04,Li_AM02,Li_APL01} and random defect effects \cite{SemenovskayaJAP1998,Semenovskaya_Ferroelectrics98}. We describe the various aspects of the theory only briefly here. For a comprehensive review of the theory, see pioneering work by L. Q. Chen et al \cite{lqchen_JACS08}.  
%,Li_PRB06,Li_APL06,Li_JAP05,chowdhury_JACE05,tenne_PRB04,Li_AM02,Li_APL01,Semenovskaya_Ferroelectrics98,SemenovskayaJAP1998,chowdhury_APL07,rao_APL07,sohJACE06,zhangAM05,ahlualiaPRB05,WangAM2004}.

The relevant contributors to the thermodynamic energy of the ferroelectric system are the bulk energy, electrostatic energy, elastic energy and the domain wall energy. The energy gained due to the phase transition from the paraelectric to the ferroelectric phase in a homogeneous unstrained ferroelectric is called the bulk energy and parameterized using the Landau coefficients. The bulk energy is given by
%\begin{equation}
\begin{align}
F_{bulk}=\alpha_{1} &(P_{x}^{2}+P_{y}^{2}+P_{z}^{2}) + \alpha_{11} (P_{x}^{4}+P_{y}^{4}+P_{z}^{4})\nonumber\\+&\alpha_{12} (P_{x}^{2}P_{y}^{2}+P_{y}^{2}P_{z}^{2}+P_{z}^{2}P_{x}^{2})
\end{align}
%\end{equation}

Here, P$_{i}$ are the polarization along the three (001) directions of the BFO crystal. The $\alpha$'s are the relevant Landau coefficients for different ferroelectrics. Note also that henceforth we will use $i$ to denote crystallographic directions and $I$ to denote the lab coordinate axis along which we will set up our numerical grid. Space directions are denoted by $x$ for crystal directions and $X$ for grid directions. 

In finite ferroelectrics, electrostatic compatibility is obtained by breaking the film into ferroelectric domains. The variation of order parameter at the domain wall causes an energy cost that originates due to both strain and dipole-dipole interaction. This additional price in energy is incorporated by the gradient of the polarization at the domain wall. The energy and the thermodynamic forces due to the domain walls are given by    
%\begin{eqnarray}
\begin{align}
F_{grad}=&G_{11} (P_{X,X}^{'}+P_{X,Y}^{'}+P_{X,Z}^{'}+P_{Y,X}^{'}\nonumber\\&+P_{Y,Y}^{'}+P_{Y,Z}^{'} +P_{Z,X}^{'}+P_{Z,Y}^{'}+P_{Z,Z}^{'})
\end{align}
\begin{equation}
\frac{\delta F_{grad}}{\delta P_{i}}=G_{11} \nabla^{2}P_{i} 
\end{equation}
%\end{eqnarray}
Here, $P^{'}_{I,I}$ are the gradient of polarization along the grid directions. $G_{11}$ is the domain wall energy coefficient when the grid and crystal directions coincide. We only consider the first order term here. $P_{I}$ is the projection of the polarization along the grid direction $I$. 

In the phase field description, the inhomogeneous long range electrostatic interaction is taken into account by solving the 3D poisson equation with appropriate boundary condition. The electrostatic energy and field are given by 
\begin{eqnarray}
F_{elec}=-\vec E\cdot \vec P\\
-\frac{\delta F_{elec}}{\delta \vec P_{I}}= \vec E_{I} 
\end{eqnarray}
Here, E$_{I}$ are the effective electric field along the grid directions. This field incorporates contributions due to inhomogeneous polarization. The contribution of the applied field is added to the electrostatic potential assuming a linear dielectric. The depolarization field is calculated separately by summing over all the local contributions to the global polarization.

The substrate constraint and domain variations within the ferroelectric film causes both homogeneous and inhomogeneous strain in the film. The effect of the elastic compatibility on the domain morphology is calculated by solving the stress-strain relation with thin film boundary condition. The elastic energy density is given by \cite{lqchen_JACS08}
\begin{align}
F_{elas}&=-\tiny{\frac{1}{2}} c_{ijkl}e_{ij}e_{kl}\nonumber\\
&=c_{ijkl} (\epsilon_{ij} - \epsilon_{ij}^{0})( \epsilon_{kl}-\epsilon_{kl}^{0})
\end{align}
Here, c$_{ijkl}$ is the elastic modulus,  $\epsilon_{ij}$ is the total strain and $\epsilon_{ij}^{0}$ is the eigen strain and $e_{ij}$ is the elastic strain.

Once the above thermodynamic energy contributions are included in the total energy expression, the thermodynamic driving force is calculated as the derivative of the total energy of the system with respect to the polarization. The subsequent temporal evolution of the polarization in 3 dimensions is calculated by using the time dependent Ginzberg- Landau equation. 
\begin{align}
\frac{\delta \vec P}{\delta t}=&-\frac{\delta F}{\delta \vec P}+\zeta_{i}(\vec r,t)\nonumber\\
=-&\frac{\delta F_{bulk}+\delta F_{wall}+\delta F_{elec}+\delta F_{elas}}{\delta \vec P}+\zeta_{i}(\vec r,t)
\end{align}
Here, $\zeta_{i}$($\vec r$,t) is the random force due to thermal fluctuation that has a zero mean and a gaussian variance.

The bulk energy and the domain wall energy are local interactions and hence are candidate for direct parallelization over the distributed processors. The nonlocal interactions on the other hand are difficult to parallelize. The non-local interaction in ferroelectrics arise due to the inhomogeneous electrostatic and elastic field. The computation of these interactions will be described in the Numerical Implementation section.
   
\section{NUMERICAL IMPLEMENTATION} 
We have implemented a time domain phase field model contrary to that described in\cite{lqchen_CPC1998}. In general, the semi-implicit fourier-spectral method allows one to take significantly longer time steps compared to a Forward-Euler method \cite{lqchen_CPC1998}. However, we employed a velocity verlet method that allowed us to take time steps significantly longer than the Forward-Euler method and reproduce the results predicted by the semi-implicit method \cite{lqchen_CPC1998}. Our specific motivation for pursuing a time domain implementation is to take advantage of modern distributed computing architectures. We will show below that problem sizes of N=10$^{9}$ can be modeled very efficiently with a time domain implementation exploiting the parallelization achieved at every computation steps. This is a significant step forward in terms of numerical capability compared to the state of the art phase-field modeling. Also, a time domain approach allows for easy and intuitive incorporation of electrostatic and mechanical boundary conditions and therefore predictive simulation of dynamic behavior can be performed. In our model, the simulation grid consists of FEM and FD grid for elastic and electrostatic calculations respectively. Below we describe the various aspects of the numerical implementation.

\begin{figurehere}
\includegraphics[width=0.9\columnwidth]{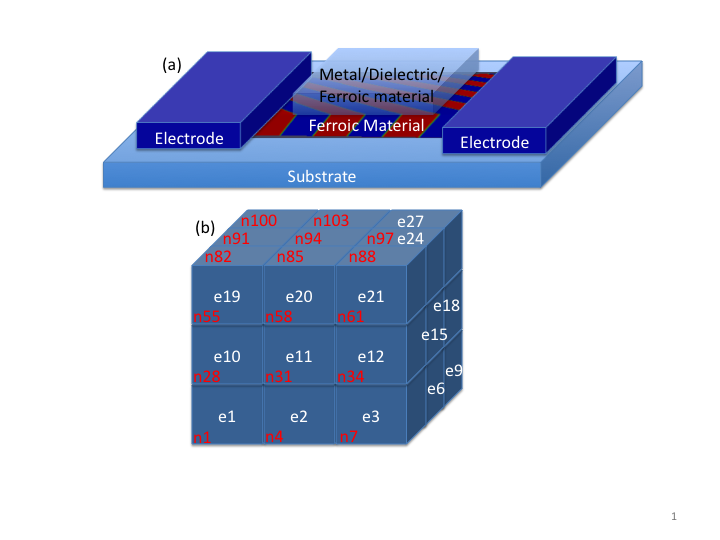}
\caption{\label{fig:Fig1}(a) Schematic of a representative device structure that is used as a test case for the developed method. Here, a thin film of ferroelectric material is grown on a substrate. Two electrodes are placed in order to apply an in-plane electric field. The electrostatic boundary condition on the material can be changed by using different materials on the ferroic thin film. The substrate strain can be varied by changing the substrate material with different lattice vectors. (b) The implemented numerical grid that contains both finite difference and finite element grids. The elements are a small block of linear brick element. The nodes of the block coincide with the FD grid. Both FEM and the FD grids are numbered in natural ordering. }
\end{figurehere}

\subsection{Grid}  
We used a mixed grid for implementing the electrostatic and elastic interactions. The elastic interaction calculation is done on a FEM grid and the electrostatic calculation is performed on a FD grid. The relationship between the FEM element and FD node numbering are shown in Fig. \ref{fig:Fig1}(b). Initially, polarizations are defined on a 3D FD grid. The bulk energy and domain wall energy are calculated on this grid. The nonlocal electrostatic field is calculated by solving the Poisson equation on the FD grid. The matrix was parallelized for a 3D FD stencil, so that maximum of the connected grid points are on the same processor. This way of parallelizing the FD stiffness matrix provides very fast matrix assembly performance and is implemented in the DA data structure of PETSC\cite{petsc-web-page}. The spontaneous strain at every node is calculated from the polarizations [see equation 16, to be discussed later]. The body force due to the spontaneous strain is assigned at the nodes of the FEM grid element nodes. A finite element grid with linear brick element is then used to solve for the stress-strain relationship. Once the total strain is calculated by solving the stress-strain relation with FEM, it is assigned in a reverse manner to the respective FD node points. The change in free energy is calculated due to the strain and hence the thermodynamic force due to the elastic energy. A natural coordinate numbering was employed for the FEM calculation. For a 3D brick element, each body node has 8 elements connected to it and hence has a total of 26 element nearest neighbor nodes. Due to a significantly increased nearest neighbor in the FEM grid, compared to the FD grid, we parallelized the stiffness matrix in the natural coordinate numbering order. Special care was taken for minimizing nonlocal assignments to the stiffness matrix when assembling, as will be described in the FEM calculation section. Thus parallelizing the two grids in two different manners facilitates maximum efficiency for the respective problems.

%\subsection{Iterative Solver}
%\textbf{NEEDS WRITING}
%We use the parallel numerical solver PETSC \cite{petsc-web-page} for solving the large set of linear equations in a parallel fashion. PETSC uses its linear solver KSP. The linear solver is used to calculate the non-local potential and the inhomogeneous strain. 

\subsection{FEM calculation}
The FEM calculation involves 1) calculating the element stiffness matrices, 2) assembling the structure stiffness matrix, 3) applying mechanical boundary conditions, 4) calculating the body forces originating from plastic strain and solve for the total strain. We describe the FEM calculation procedure below.

\subsubsection{Element Stiffness Matrix}
We used an iso-parametric linear brick element for implementing the FEM calculation of the inhomogeneous strain. An iso-parametric implementation of the elements allows for Gaussian Integration during element stiffness calculation and thus facilitates faster assembly. The procedure is as follows. First the iso-parametric element is written in natural coordinates. Then the shape function is calculated in terms of the natural coordinates of the element nodes. 
\begin{eqnarray}
\frac{\delta N^{I}}{\delta \zeta^{I}}_{8\times3}=\left[ 
\frac{\delta N^{I}}{\delta \zeta}	
\frac{\delta N^{I}}{\delta \beta}	
\frac{\delta N^{I}}{\delta \eta}
\right]
\end{eqnarray} 
Here, ${\zeta}$, ${\beta}$ and ${\eta}$ are the axes in natural coordinates. N$^{I}$ are the components of the shape function along the natural coordinates.  

The real space node locations are given by the matrix,
\begin{eqnarray}
R_{8\times3}=\left[ X_{I}	Y_{I} 	Z_{I}\right]
\end{eqnarray}
Here, $X_{I}$ are the node points of the element in real space coordinates along the grid directions. 
The Jacobian matrix is calculated as
\begin{eqnarray}
J_{3\times3}=
\left( \frac{\delta N^{I}}{\delta \zeta^{I}}
%\right)^{T} _{3\times8} \times R_{8\times3}
\right)^{T} R
\end{eqnarray}
The space derivatives of the shape function in terms of the natural coordinates are given by,
\begin{eqnarray}
\frac{\delta N^{I}}{\delta X^{I}}_{3\times8}
%&=&J^{-1}_{3\times3} \times \left(\frac{\delta N^{I}}{\delta \zeta^{I}}\right)^{T}_{3\times8}\nonumber\\
&=&J^{-1} \left(\frac{\delta N^{I}}{\delta \zeta^{I}}\right)^{T}\nonumber\\
&=&
\begin{matrix}
\bordermatrix{&
\frac{\delta N^{I}}{\delta X}\\
&\frac{\delta N^{I}}{\delta Y}\\
&\frac{\delta N^{I}}{\delta Z} }
\end{matrix}
\end{eqnarray}

The strain-displacement matrix is constructed from the shape function as 
\begin{eqnarray}
B_{6\times24}=
\begin{matrix}
\bordermatrix{&
\frac{\delta N^{I}}{\delta X} & 0 & 0\\
&0 & \frac{\delta N^{I}}{\delta Y} & 0\\
&0 & 0 & \frac{\delta N^{I}}{\delta Z} \\
&\frac{\delta N^{I}}{\delta Y} & \frac{\delta N^{I}}{\delta X} & 0\\
&0 & \frac{\delta N^{I}}{\delta Z} & \frac{\delta N^{I}}{\delta Y} \\
&\frac{\delta N^{I}}{\delta Z} & 0 & \frac{\delta N^{I}}{\delta X} }
\end{matrix}
\end{eqnarray}
The anisotropic stiffness coefficient is given by
%\begin{eqnarray}
\begin{align}
C_{6\times6}&= \frac{E}{(1+\gamma)(1-2*\gamma)}\times \nonumber\\
&\begin{matrix}
\bordermatrix{
&1-\gamma & \gamma & \gamma &  & & \\
&\gamma & 1-\gamma & \gamma &  &  & \\
&\gamma & \gamma & 1-\gamma &  & & \\
&  & &  &\frac{1-2*\gamma}{2} & & \\
&  & &  & &\frac{1-2*\gamma}{2} & \\
&  & &  & & &\frac{1-2*\gamma}{2} }
\end{matrix}
\end{align}
%\end{eqnarray}
Here, E is the elastic stiffness coefficient and $\gamma$ is the Poisson ratio. 
%\end{eqnarray}

The element stiffness matrix is K$^{e}$ and is calculated by Gaussian integration in the natural coordinates. 
%\begin{eqnarray}
\begin{align}
K^{e}_{24\times24}&=\int_{-c}^{\:c}\int_{-b}^{\:b}\int_{-a}^{\:a}B(R^{I})^{T}CB(R^{I}) dX dY dZ\nonumber\\
&=\int_{-1}^{\:1}\int_{-1}^{\:1}\int_{-1}^{\:1}\mid J\mid B(\zeta^{I})^{T}CB(\zeta^{I}) d\zeta d\beta d\eta \nonumber\\
&=\sum_{\zeta=\frac{\pm1}{\sqrt{3}}}\sum_{\beta=\frac{\pm1}{\sqrt{3}}}\sum_{\eta=\frac{\pm1}{\sqrt{3}}} \mid J\mid B(\zeta^{I})^{T}CB(\zeta^{I}) 
\end{align}
%\end{eqnarray}
Here, $\pm$(a,b,c) are the node point coordinates of the linear brick element in real space. 
\begin{figurehere}
\includegraphics[width=0.9\columnwidth]{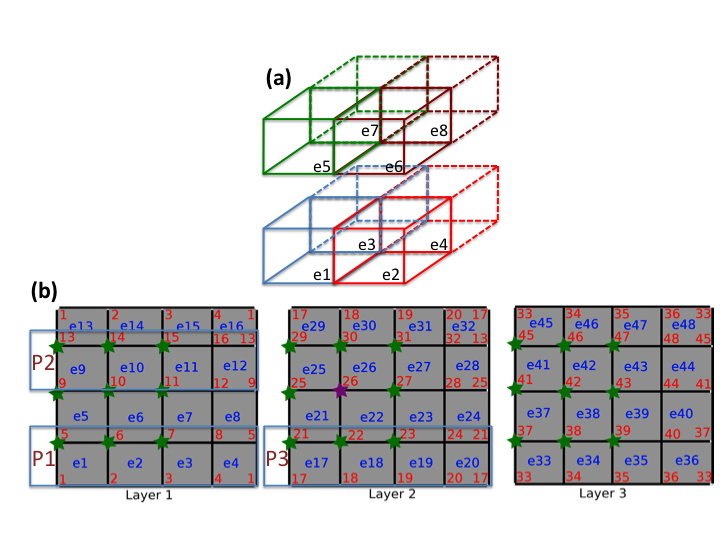}
\caption{\label{fig:Fig2}(a) FEM structure assembly by element. The newly added element (e8) nodes have matrix element contributions from the 7 elements (e1-7). Note that the contributing elements are only those that precede this element in the natural grid along the three directions. (b) FEM structure matrix assembly by node. The node in consideration is 26 (violet star). Due to the element connectivity the node has interaction with its in-plane surrounding nodes and also the layers above and below this node (green stars). Each node has a total of 26 connected nodes within the body of the structure. The number of connected element and nodes vary at the boundary. These boundary nodes and elements are assembled in a similar procedure with appropriate connectivity. \\}
\end{figurehere}

\subsubsection{Assembly}
\label{subsec:assembly}
Once the element stiffness matrices are calculated for each of the element, the structure stiffness matrix is constructed by assembling the element stiffness matrices. Assembling the structure stiffness matrix is the main bottleneck in achieving high performance in parallel computation. We used a natural coordinate numbering for the FEM grid (X varying fastest, then Y and then Z). Thus element nodes varying along X direction are nearest neighbor in the global node numbering. In this grid numbering method, the FEM stencil for linear brick elements are widely separated on distributed processors for large number of degrees of freedom (DOF)  as shown in Fig\ref{fig:Fig2}. The element connectivity is shown in Fig\ref{fig:Fig2}(a). Here, the added element 8 is connected to other 7 elements that are prior to this element in grid numbering system. Note that the connectivity is only backward, meaning the element stiffness matrix of e8 only depends on the nearest neighbors e1-e7. Elements that are added after e8 to the global grid are not essential for calculating the stiffness matrix of e8. Since each element spans 2 layers, any node has nearest neighbors in the adjacent layers in all the directions. The node connectivity for the FEM stencil is shown in Fig \ref{fig:Fig2}(b). A specific node under consideration is labeled 26 with a violet star sign. It has 26 connected nodes in 8 elements that contribute to different columns of the row associated with this node. This type of multilayer connectivity of elements makes it very difficult to assemble the structure matrix avoiding non-local assignments. In fact, with our model of element wise assembly process, even a 32$\times$32$\times$3 elements global grid required about 120 seconds to assemble on moderate 4 processors. The poor performance of the element wise assembly originates from the fact that different nodes associated with the nearest neighbor element belong to different processors (Fig \ref{fig:Fig2}(b)). The communication burden between processors overwhelms the computation benefit even for a small global grid size. Hence, we discarded the element wise assembly process and resorted to a node wise assembly process. The essential idea is to reduce the processor to processor communication as much as possible during global stiffness matrix assembly even at the cost of increased computation within individual processors. We calculate the element matrix elements for eight elements that are connected to a specific node locally. Theoretically this amounts to calculating the matrix element for each element 8 times assuming the worst case scenario where each node of an element belongs to a different processor. However practically, for large global matrices with natural coordinate numbering, the 8 nodes of an element belong to only 2 processors. The 8 nodes are divided into bottom and top layers, each containing 4 nodes and owned by individual processors. This amounts to calculation of the element stiffness matrices only twice instead of eight times. However, even with this double calculation, the node wise assembly process makes the whole element stiffness matrix calculation and assembly local to individual processors. Thus linear scaling performance can be achieved in the global matrix calculation and assembly process for arbitrarily large structures. Note that in comparison, the element wise assembly shows poor scaling performance even for 4 processors for a moderately small grid size. The algorithm for implementing the node wise assembly is as follows:
i) Determine the nodes owned by each processor and iterate through them.
ii) At each iteration, determine the global nodes associated with the 8 nearest neighbor elements  of the node and write the global indices of the 27 nearest neighbor nodes in an (27$\times$ 3) array. This array contains the column numbers of the global matrix
where the matrix elements will be written.   
iii) At each iteration, determine the global element number of the 8 elements associated with the node and iterate through them
iv) For each element determine to what node of the element is the node of interest (26) connected   
v) Iterate through the eight nodes of the element
vi) Fetch the values from the row of element matrix that corresponds to the node number determined in step iv and add the values. 
   
Thus all the assignments in constructing the global matrix are local. In this manner, we avoid nonlocal assignment of adding values to the structural matrix and achieve more than two orders of magnitude boost in assembly process even for moderate size of the grid(32$\times$32$\times$32$\times$3) compared to direct calculation of element matrices. 

\subsubsection{Body Force}
The polarization causes an eigen strain in the unit cell. The eigen strain is included in the elastic equation as a body force. First, the eigen strain is calculated using the electrostriction coefficients. The body forces due to the eigen strain for each element are calculated and assigned to the element nodes. The eigen strain and the body forces are given by 
%\begin{eqnarray}
\begin{align}
\epsilon ^{0}_{ii}&=Q_{11}P_{i}^{2}+Q_{12}(P_{j}^{2}+P_{k}^{2}) \nonumber\\
\epsilon ^{0}_{ij}&=Q_{44}P_{i}P_{j} \nonumber\\
F_{i}^{0}&=\int_{\Omega}[B]^{T}[C]{\epsilon ^{0}_{kl}} d\Omega
%\end{eqnarray} 
\end{align}
Here, $Q_{ij}$ are the electrostriction coefficients. 

Next, the elastic equilibrium is calculated from
\begin{eqnarray}
Kdu_{i}=F_{i}^{0}
\end{eqnarray}   
Here, $K$ is the structure stiffness matrix including the appropriate boundary condition and $du_{i}$ is the displacement.

\subsubsection{Boundary Conditions}
The boundary conditions are applied by making relevant changes to the matrix that contains information for structure stiffness. For example, in the thin film structure, the regularly used boundary condition is periodic along $XY$, clamped at the bottom and stress free at the top. The periodic boundary condition can be applied during assembly process assuming the elements at the left boundary to be the neighbor for elements on the right boundary. The clamped boundary condition is applied by zeroing all the values in the matrix corresponding to the bottom interface leaving the diagonal entries as 1. A specific displacement can be assigned to these nodes by setting these values on the right hand side. The stress free boundary condition is similarly applied by zeroing the forces at the corresponding nodes along the appropriate direction.

Once these modifications due to the boundary conditions are employed to the matrix and the right hand side vector, the resulting total displacement is calculated by solving the $Kdu=f$ equation. Again we use an iterative Krylov subspace solver for solving the set of linear equations.

\subsection{Electrostatics Calculation}
Electric field due to the inhomogeneous polarization and the applied field incorporating the appropriate boundary condition is calculated on the FD grid. The charge density resulting from the inhomogeneous polarization and the resulting potential are calculated using 
\begin{eqnarray}
\rho=\nabla \cdot P\\
\epsilon \nabla^{2} \phi=\rho
\end{eqnarray} 
Different boundary conditions can be easily applied when assembling the laplacian operator in 3D. Some of the useful boundary conditions are periodic, Dirichlet, Neumann and inhomogeneous material interface. The applied field is incorporated by assigning a predefined value that is a Dirichlet boundary condition where the electrodes are placed. The floating boundary or the Neumann boundary condition is applied on open surfaces by setting the normal component of the electric field to zero.

\subsection{Time Integration}
We tested both the explicit forward-euler and the velocity verlet algorithm. The latter algorithm allowed for an order of magnitude longer time steps compared to the direct forward euler method for stiff problems. In order to check the accuracy of the results, we simulated the results presented in Ref. \cite{lqchen_CPC1998}. Both equilibrium and kinetic results were tested. With reduced space discretization length of 1nm and reduced time step size of 0.01, we reproduced the results presented in Ref. \cite{lqchen_CPC1998}.  We used the same space and time discritization length for the results reported here. The velocity-verlet integration algorithm implemented is given by the equations. 
%\begin{eqnarray}
\begin{align}
P^{p}_{i+1}&=P^{c}_{i}+\frac{\delta P^{c}_{i}}{\delta t} \Delta t \nonumber\\ 
\frac{\delta P^{c}_{i+1}}{\delta t}&=\frac{1}{2}\left(\frac{\delta P^{c}_{i}}{\delta t}+\frac{\delta P^{p}_{i+1}}{\delta t}\right) \nonumber\\
P^{p}_{i+1}&=P^{c}_{i}+\frac{\delta P^{c}_{i+1}}{\delta t} \Delta t 
\end{align}
%\end{eqnarray}

\begin{figurehere}
\includegraphics[width=0.9\columnwidth]{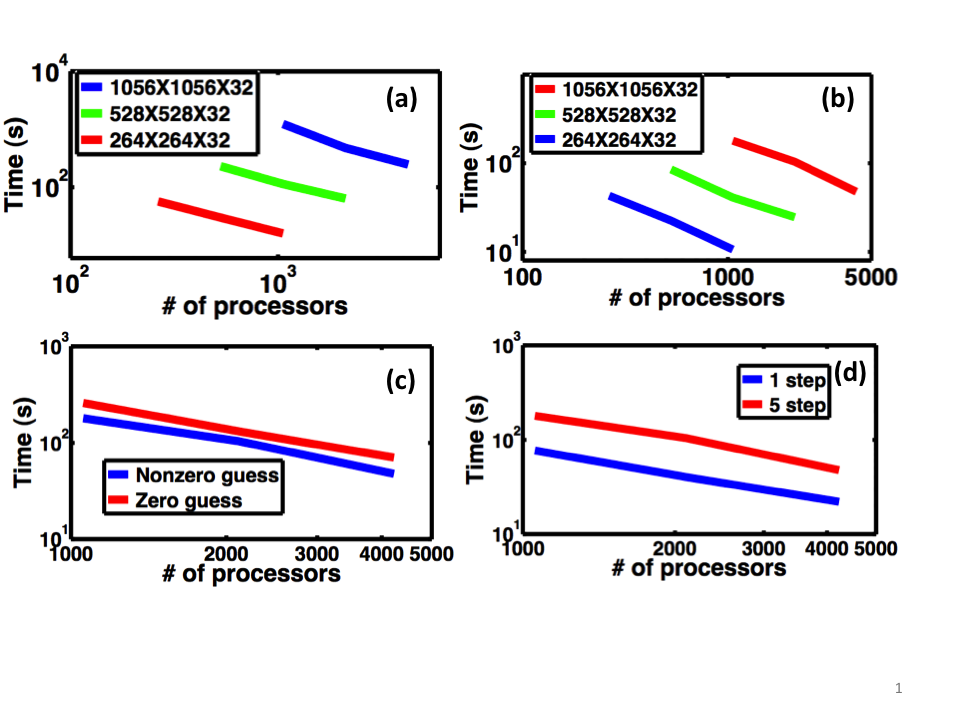}
\caption{\label{fig:Fig3}(a) The FEM stiffness matrix assembly time as a function of the number of processors used with varying grid size. (b) The total cycle time for 6 iterations including the nonlocal electrostatic and elastic interactions as a function of the number of processors used with varying grid size. (c) Using a non-zero initial guess from the last time step solution, for the linear solver during the electrostatic and elastic interaction calculation improves the overall performance of the  by a factor of 2 for all number of processors used. (d) Calculation of the long range interactions every 5 steps compared to every single step, improved the performance by about 4 times. This does not change the physical results since, the long range interactions usually act at low frequency compared to the short range interaction. For all the three structure sizes, we obtain linear scaling.\\}
\end{figurehere} 

\subsection{Numerical Performance}
The parallel performance of the 3D phase-field code was analyzed on Hopper of the NERSC facility. The Hopper machine is equipped with 24 2.1 GHz processors and 32 GB memory per node. It uses Gemini interconnect for inter node communication that has a latency of $\sim$1 ${\mu}$s. In FIG \ref{fig:Fig3}(a) we show the parallel performance of the FEM assembly algorithm. Three different sizes of the grid are assembled from a micron to a quarter micron square device. The smallest structure is assembled using 264 processors taking 57 seconds. When 1056 processors are used, the assembly time reduces to 16 seconds. Thus we achieve a nearly linear scaling for the assembly process. Similar scaling behavior is also achieved for larger structures (up to micron size) using more number of processors (up to ~5k) as shown in FIG \ref{fig:Fig3}(a). For small deformations of the structures during the switching process, it is sufficient to work with the initial FEM stiffness matrix at subsequent time steps. Hence, we separate the assembly process and cycle time for each iteration of the self-consistent phase-field calculation. Each time step of the phase-field calculation consists of calculating the local bulk energy, domain wall energy and the non-local electrostatic and elastic energy. In FIG \ref{fig:Fig3}(b), we show the total cycle time of 6 iterations. For the smallest grid (same as FIG \ref{fig:Fig3}(a)), the average time for a single time step is ~7.1 s with 264 processors. The cycle time reduces to ~1.7 s when 1056 processors are used. Again, the computation time per cycle scales linearly with increasing number of processors. The larger grid sizes with increasing number of processors show similar scaling behavior. For calculation of non-local interactions during time stepping, if we initialize the linear solver with the result of the last time step, the number of iteration required for the linear solver used in both electrostatic and elastic calculation, decreases significantly. We gain an overall two fold performance benefit for all structure sizes tested across different number of processors as shown in FIG. \ref{fig:Fig3}(c)). We also found that, when simulating the spatially extended dynamics of a ferro-electric structure, the short range interactions gives rise to stiffness. However, the long range interactions usually have a longer temporal wavelength. Hence, for fixed size of the time step, it is reasonable to sample the long range interactions every few sample of the short range dynamics. Hence, we calculate the long range interaction once after every 5 steps of the short range interaction calculation. We find that this approximation does not change the physical results obtained. However, the overall cycle time reduces by about 4 fold as shown in FIG. \ref{fig:Fig3}(d). Thus significant improvement in scale can be achieved by a parallel implementation of the phase-field model. Reaching the micron scale could enable direct comparison of simulation results with experimental observation that we discuss next.

\section{RESULTS}
\label{sec:RESULTS}

%\noindent .\\

In this section, we show the calculation results of ferroelectric switching and domain pattern evolution on the 001 surface of BFO and emphasize the importance of electrical boundary condition on the observed switching pattern. The presented simulation results have the dimension of 1056nm$\times$1056nm$\times$32nm. The energy was normalized so that dimensionless space-time are obtained following Ref. \cite{Hu_MSEA1997}. The spatial grid size is 1 nm and time step size is 0.005. The thermodynamic parameters for BFO were obtained from Ref. \cite{zhang_JAP07}. A generic device structure incorporating the multiferroic material is shown in FIG. \ref{fig:Fig1}(a). The device shows the ferroic material is coherently strained by the substrate. An in-plane electric field is applied using the two in-plane electrodes. The electrostatic boundary condition of the film surface is controlled by placing a metal or dielectric or simply leaving it open. This representative device structure is a prototype of many recent experimentally reported ferroelectric devices.

\begin{figurehere}
\includegraphics[width=0.9\columnwidth]{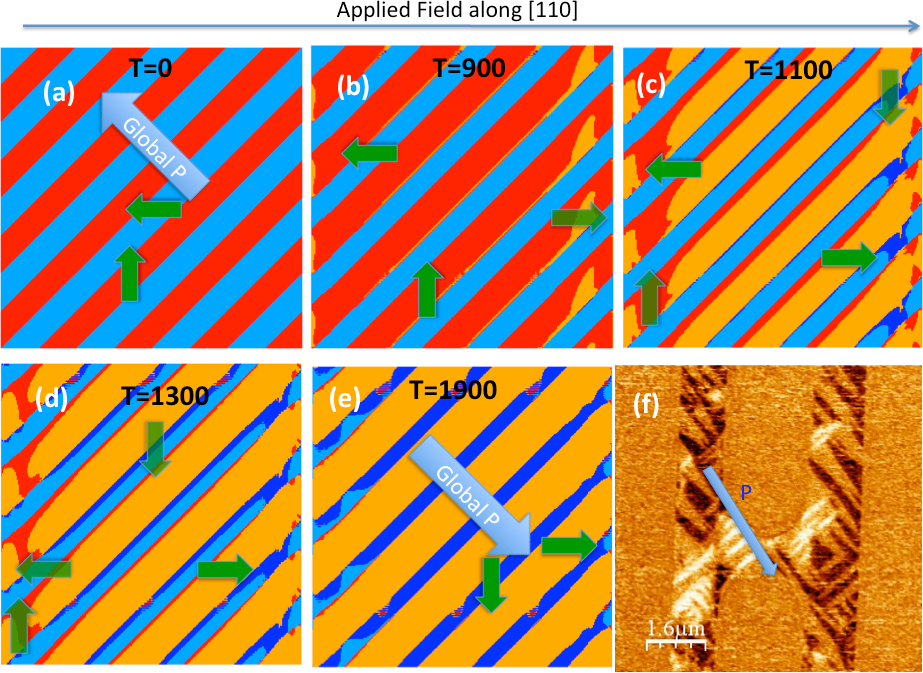}
\caption{Evolution of the polarization on the (001) surface under short circuit boundary condition. (a) The initial domain pattern with left (light blue) and up(red) polarization domains. (b) Nucleation of right polarization domain (yellow) through 71$^{\circ}$ switching of up domains. (c) 71$^{\circ}$ switched domain (yellow) grows. (d,e) A new domain grows towards south (deep blue) and eventually switches the whole domain. The global polarization switches by 180$^{\circ}$ in the process. (f) Experimental observation of 180$^{\circ}$ switch of the domains under a short circuit boundary condition (from Ref. \cite{john_PRL11})}.
\label{fig:Fig4}
\end{figurehere} 

In this particular case, a metallic boundary condition was applied on the top (001) surface. We assumed that the thin film is coherently strained by the substrate. A low electric field was applied along the [110] direction that is just above the coercive field of the 71$^{\circ}$ switch. The evolution of the thin film BFO as a function of time is shown in FIG. \ref{fig:Fig4}. The application of an in-plane voltage on a BFO thin film with striped domain structure can only generate in-plane 71$^{\circ}$ and 109$^{\circ}$ switching events. BFO's thermodynamic potential profile is such that E$_c$(71$^{\circ}$) $<$ E$_c($109$^{\circ}$). The calculated coercive field of the 71$^{\circ}$ switch is 420 kV/cm and for the 109$^{\circ}$ switch it is 490 kV/cm. Considering the as grown 71$^{\circ}$ striped BFO configuration represented in FIG. \ref{fig:Fig4}, the high saturation polarization ($~$90 $\mu$C/cm$^{2}$) of BFO causes all of the domains to arrange in-plane in a head-to-tail configuration so that the dipole-dipole energy is minimized. Now, for an applied electric field of strength E$_c($109$^{\circ}$) $<$ E$_{applied}$ $<$ E$_c$(71$^{\circ}$) and directed from left to right (here along [110] BFO), our calculation demonstrates that the ferroelectric domains with an in-plane polarization oriented perpendicular to the applied electric field align first towards the direction of this external field (from [\=11\=1] BFO to [11\=1] BFO )(Fig. 4(c)). This corresponds to a 71$^{\circ}$ switching event due to the applied field (in-plane switching of the ferroelectric domain oriented antiparallel to the electric field would correspond to a 109$^{\circ}$ switching event, from [\=1\=1\=1] BFO to [11\=1] BFO). The ferroelectric switching begins at the domain wall. Although, the simulation was started from a uniform polarization distribution within the domains, some domain regions(yellow) switch earlier than the rest of the domains. This specific pattern formation during the switching process is a result of the long range electrostatic interaction. Non-uniform charge originates in the domain during switching which causes this electrostatic field. Thus some region of the domain are under a higher effective electric field compared to the other regions. Eventually the whole domain switches along the applied field direction. The switched domain (yellow) generates an energetically unfavorable head-to-head configuration at the wall. The theoretical maximum limit of the dipole-dipole fields at these domain walls can reach up to ~10$^{4}$ kV/cm due to the induced charge. Domains originally oriented antiparallel to the electric field (along [\=1\=1\=1]BFO) then switch in-plane by 90$^{\circ}$ (corresponding to a second 71$^{\circ}$ switching event to the direction [1\=1\=1] BFO ) under this dipole-dipole field to recover the preferred head-to-tail configuration of the polarizations (dark blue region). The growth of a new domain also exhibits a pattern due to electrostatic interaction. Note that, this long range electrostatic interaction is in-plane due to the metallic boundary condition applied on the top surface. We have described a second order switching that occurs in this configuration and compared to experiment in ref. \cite{john_PRL11}.

%\vspace{-2.5cm}
\begin{figurehere}
\includegraphics[width=0.9\columnwidth]{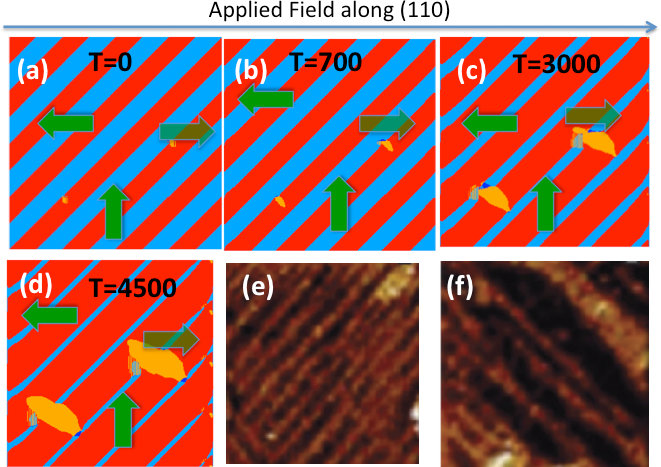}
\caption{\label{fig:Fig5}Evolution of the polarization on the (001) surface with an open boundary condition. (a) The initial domain pattern with left (light blue) and up(red) oriented polarization domains. (b) Anisotropic growth of right oriented domain(yellow) through a 71$^{\circ}$ switch of the up(red) oriented polarization to right oriented domain(yellow). (c) The up(red) oriented domain grows simultaneously through domain wall switching of the left (light blue) domain. (d) Emergence of domain patterns (between red and yellow domains) aligned at 90$^{\circ}$ to the initial domain pattern. (e,f) PFM image showing the 90$^{\circ}$ switch of domain pattern under open circuit boundary condition (from Ref. \cite{chu_NatMat08}) \\}
\end{figurehere}

In order to emphasize the important role that electrical boundary conditions play for the domain pattern reorganization during lateral electrical switching, we simulated the same device structure with open boundary condition. A Neumann boundary condition was applied on the top surface with the top 5 layers as air. Due to the difference in dielectric constant between BFO ($\epsilon$=100) and air ($\epsilon$=1), most of the electric field go through air. The 3D electrostatic interaction makes the growth of the domains anisotropic. When an open circuit boundary condition is applied, a completely different result in terms of the domain pattern is obtained as shown in FIG. \ref{fig:Fig5}. We introduced two nucleation centers in the initial domain in order to study how these nucleated domains grow when an electric field is applied. These switched regions (yellow) cause a charged domain walls with the initial domain (red). The charges at the domain wall is reduced by growth of the switched domain(yellow) perpendicular to the initial domain wall as shown in FIG. \ref{fig:Fig5}(b). When the growing domain (yellow) enters the left oriented (light blue) region (FIG. \ref{fig:Fig5}(c)), the polarizations in the two domains (light blue and yellow) are head to head. In this region, the yellow domain grow perpendicular to the electrodes rather than perpendicular to the initial domains. However, due to domain wall switching, most of the region is up directed (red). Hence, the growth of the newly formed domain(yellow) is mostly perpendicular to the initial domain wall. Overall, we observe a new domain pattern consisting of red and yellow regions that is at 90$^{\circ}$ to the initial domain pattern as shown in FIG. \ref{fig:Fig5}(d). Experimentally this type of switching has been observed in planar switching of BFO with open circuit boundary condition \cite{chu_NatMat08}. Note that the domain walls have reconstructed and  the width of the domains with polarization along the applied field is higher. This 90$^{\circ}$ reorientation of the domain pattern with an applied field is solely due to the anisotropic growth of domains that occur at the domain walls. 

\begin{figurehere}
\includegraphics[width=0.9\columnwidth]{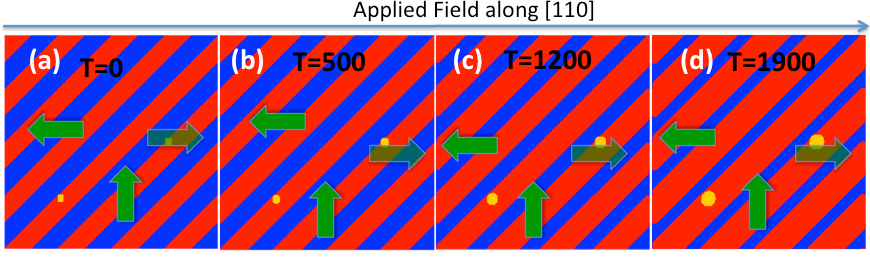}
\caption{\label{fig:Fig6}Evolution of the polarization on the (001) surface with an open boundary condition without considering the domain wall charge. (a) The initial domain pattern. (b) Isotropic growth of right oriented polarization domain (yellow) through an 71$^{\circ}$ switch of the up polarization (red). (c,d) Gradual isotropic growth of the switched domain (yellow) due to the applied field. The emergent domains that do not have a specific stripe like pattern since the effect of charge was ignored. \\}
\end{figurehere} 

If we ignore the charges at the domain walls, then the individual domains grow isotropically as shown in Fig \ref{fig:Fig6}.  Due to the isotropic growth, the reconstructed pattern do not show any specific topography also the domain wall speed is significantly slow. This result is unphysical, since the presence of electrical charge should make the pattern stripe like.   

\vspace{-0.9cm}
\subsection{Applied Field Along [100] Direction}
Study of temporal evolution of ferroelectric domain switching is facilitated by recent advent of in-plane capacitor geometry and PFM analysis. There are a number of recent experimental report on the temporal evolution of the ferroelectric domain under a lateral applied electric field \cite{you_planarBFOSwitch_APL10,balke_BFOplanarSwitch_AM2010}. Here, we show the simulation result of ferroelectric domain switching on the (001) plane of BFO when the applied field is along the [100] direction, as shown in Fig \ref{fig:Fig7}. An open circuit boundary condition was applied in the same manner as described in the previous section. The device geometry corresponds to that of \cite{you_planarBFOSwitch_APL10}. Similar switching mechanism applies to that reported in \cite{balke_BFOplanarSwitch_AM2010}. The simulated device dimension is 1056nm$\times$264nm$\times$32nm. 

\begin{figurehere}
\includegraphics[width=0.9\columnwidth]{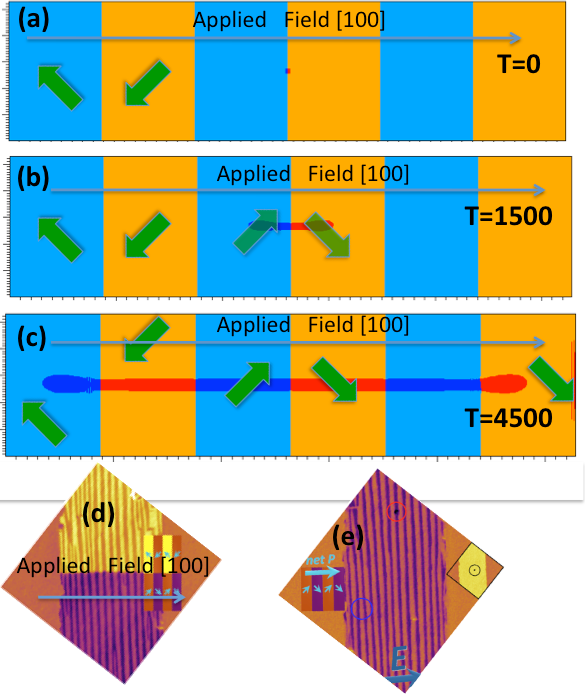}
\caption{\label{fig:Fig7}Evolution of the polarization on the (001) surface with an open boundary condition when a field is applied along the [100] direction. (a) The initial domain pattern with a defect introduced where the switching starts(dark blue and red dot). (b) Anisotropic growth of right oriented polarization domain (dark blue and red) through an 71$^{\circ}$ switch of the left polarization (light blue and yellow) along the applied field direction. (c) The anisotropic growth continues and switches regions close to the electrode. Slow growth perpendicular to the applied field and retention of the domain size matches very well with the experimental observation. (d,e) Experimental data showing the intermediate stage between switching(from Ref. \cite{you_planarBFOSwitch_APL10}). \\}
\end{figurehere} 

There are two initial domain variants on the (001) surface. The [\=11\=1] (light blue) and [\=1\=1\=1] (yellow). A single nucleation site was introduced as a nucleation point in the [\=11\=1] domain where the nucleated polarization is oriented along the [11\=1] direction. A voltage of -75 V was applied on the right electrode while the left electrode was grounded. In this case, the switching initially occurs at the nucleation center due to strained walls where the [\=11\=1] (light blue) polarizations switch along the [11\=1] (dark blue) direction and [\=1\=1\=1] (yellow) domains switch along [1\=1\=1] (red). The initial switching is isotropic as shown in Fig. \ref{fig:Fig7}(a). However, as the switching domain grow, the walls that are parallel to the electrodes are charged (due to head to head polarization configuration). On the other hand, the walls perpendicular to the electrodes are uncharged but strained. In subsequent switching steps, the charged domain wall grows significantly faster than the sidewise strained wall. The incorporation of inhomogeneous electric field in the model capture this physical process that occur due to the creation of charged domain walls. The inhomogeneous electric fields due to the domain wall charges terminate at the electrode. The sign of the charges are such that they aid the applied field along the applied field direction. Hence the polarizations are under a higher effective field along the applied field direction in the regions where striped domain region is formed. Anisotropic domain growth occurs due to this effective electric field as shown in Fig. \ref{fig:Fig7}(b). Once a narrow stripe has been created due to forward domain growth, the domain walls perpendicular to the electrodes are charge compensated. The wall growth velocity is significantly slower in the sidewise direction than in the forward direction. At this stage, domains grow only perpendicular to the electrodes(\ref{fig:Fig7}(c)).

\begin{figurehere}
\includegraphics[width=0.9\columnwidth]{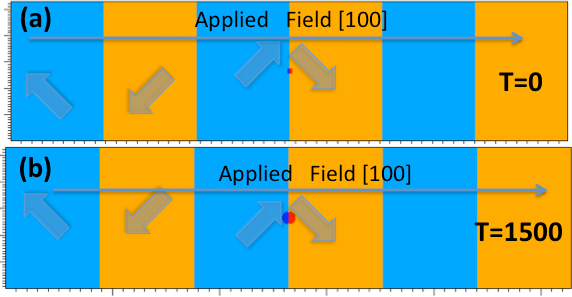}
\caption{\label{fig:Fig8}Evolution of the polarization on the (001) surface with an open boundary condition when a field is applied along the [100] direction without considering the domain wall charge. (a) The initial domain pattern with a defect introduced where the switching starts. (b) Isotropic growth of right oriented polarization domain (dark blue) through a 71$^{\circ}$ switch of the left polarization (light blue) along the applied field direction. \\}
\end{figurehere} 

The anisotropic growth of the domain wall occurs due to the creation of charge at the switched domain wall. If we do not consider the domain wall charge, then the domain growth is isotropic and a circular domain is created due to the isotropic domain wall energy. The domain growth in this case is shown in Fig. \ref{fig:Fig8}. 

\subsection{ Domain Reconstruction}

One interesting aspect of the anisotropic switching process is that in this case no domain wall reconstruction takes place. The coercive field for switching at the domain wall is in fact same as that for the strained sidewise domain switching. However, the nature of the domain wall switch is that at the domain wall, polarization from one domain switches towards the polarization in the other domain. For an applied field that is perfectly symmetric with respect to the two polarization domains, there occurs a frustrated condition for switching from initial to the final polarization direction as the probability of switching in either directions are the same. Thus we find that for an applied field that is orthogonal to the domain walls cannot move the walls during switching and no domain size reconstruction takes place. On the contrary, when the applied field is not perpendicular to the domain wall, the applied field is not symmetric with respect to the polarizations in the adjacent domains. One direction of the DW switching becomes preferable compared to the other. The DW propagates deterministically in a specific direction and the domain size reconstructs. This type of DW switching and domain size reconstruction is shown in Fig. \ref{fig:Fig4}. It is important to note that the suppression of DW switching occurs due to the same coercive field of the two domain variants at the wall. So if the saturation polarization in the two domains are different, for example due to anisotropic strain, then the coercive field for DW switching will also be different for the two adjacent domains. In this case, an applied field perpendicular to the domain wall will also cause DW switching and consequently DW movement. However, depending on the relative magnitude of the coercive fields for the two processes, it is possible to find an angle with respect to the domain wall where the coercive field for the two processes are the same. In brief, it is possible to find a direction for the applied field such that during the switching process, the DW switching is locked and no domain size reconstruction takes place irrespective of the anisotropic strain introduced by the substrate. For real applications, domain size control could be an important design parameter.

\section{Conclusion}
\label{sec:conc}
In summary, we have reported a massively parallel time domain phase field model. We have achieved nearly linear scaling for all the steps in the calculation. The near perfect scaling allows us to simulate the switching dynamics of micron scale devices. Especially, the nonlinear response of multi-domain ferroelectric films caused by arbitrary electrostatic and mechanical boundary conditions can be easily studied and direct comparison with the experimental results can be made. Here we have shown the results of lateral electric field switching incorporating the multiferroic material BFO and compared the obtained results with multiple experimental results. We have particularly emphasized the importance of charge driven domain growth mechanism that explains the temporal evolution of ferroelectric domains observed in experiments. We have also elucidated the role of electrostatic boundary condition on the lateral multi-domain ferroelectric switching. Finally, we predict a method of domain size control during ferroelectric switching which could find important applications in device design using these materials. We believe that the model will be useful in predicting device operation at the length scale where a lot of current experiments are being performed.  

This work supported in part by Nanoelectronic Research Initiative (NRI) and National Science Foundation (NSF). The computer simulations were performed at NERSC. 

\vspace{-0.5cm}
%\newpage
%\bibliography{BIBLIOGRAPHY}
%\begin{thebibliography}{10}

%

\newpage
%
%------------------------FIGURES-----------------------------
%
\newpage

\begin{center}

\end{center}
\end{document}